# Biofluorescent Worlds – II. Biological fluorescence induced by stellar UV flares, a new temporal biosignature

Jack T. O'Malley-James[1]★ and Lisa Kaltenegger[1,2]

[1]*Carl Sagan Institute at Cornell University, Ithaca, NY 14853, USA*
[2]*Astronomy Department, Ithaca, NY 14853, USA*

## ABSTRACT

Our first targets in the search for signs of life are orbiting nearby M stars, such as the planets in the Proxima Centauri, Ross-128, LHS-1140, and TRAPPIST-1 systems. Future ground-based discoveries, and those from the TESS mission, will provide additional close-by targets. However, young M stars tend to be very active, flaring frequently and causing UV fluxes on the surfaces of HZ planets to become biologically harmful. Common UV-protection methods used by life (e.g. living underground, or underwater) would make a biosphere harder to detect. However, photoprotective biofluorescence, 'up-shifting' UV to longer, safer wavelengths, could increase a biosphere's detectability. Here we model intermittent emission at specific wavelengths in the visible spectrum caused by biofluorescence as a new temporal biosignature for planets around active M stars. We use the absorption and emission characteristics of common coral fluorescent pigments and proteins to create model spectra and colours for an Earth-like planet in such a system, accounting for different surface features, atmospheric absorption, and cloud cover. We find that for a cloud-free planet biofluorescence could induce a temporary change in brightness that is significantly higher than the reflected flux alone, causing up to two orders-of-magnitude change in planet–star contrast, compared to a non-fluorescent state, if the surface is fully covered by a highly efficient fluorescent biosphere. Hence, UV-flare induced biofluorescence presents previously unexplored possibilities for a new temporal biosignature that could be detectable by instruments like those planned for the extremely large telescope and could reveal hidden biospheres.

## 1 INTRODUCTION

M stars are the most common type of star in the galaxy and make up 75 per cent of the stars in the solar neighbourhood. They are also excellent candidates for habitable zone (HZ) terrestrial planet searches, due to the high frequency of rocky planets in the HZs of these stars (Dressing & Charbonneau 2013; Gaidos 2013).

Estimates of the occurrence rate of Earth-sized (0.4–1 $R_E$) planets in the surface liquid water zone (or so-called HZ) around cool dwarfs range between 15 per cent and 66 per cent (Traub 2011; Dressing & Charbonneau 2013; Gaidos 2013; Kopparapu 2013; Ari, Gaidos, & Wu 2015). The recently launched *Terrestrial Exoplanet Survey Satellite* (TESS) mission surveys nearby bright stars to identify transiting exoplanets, including terrestrial ones in the HZ. It is sensitive enough to identify HZ planet candidates around nearby low-mass stars ($T_{eff} \leq 4000$ K; late M and early K stars) for future ground and space-based characterization (Ricker et al. 2014; Sullivan et al. 2015). TESS is expected to find 100 s of 1.25–2 $R_E$ planets and 10 s of Earth-sized planets, with a handful (<20) of these planets in the HZ of their cool host stars (Ricker et al. 2014; Barnes, Meadows, & Evans 2015; Sullivan et al. 2015; Kaltenegger et al. 2019). This makes it likely that the first HZ planets that will be characterized will be orbiting nearby M stars.

Our nearest known HZ terrestrial planets – *Proxima-b, Ross 128b, TRAPPIST-1e*, *-f*, *-g*, *LHS 1140-b* (Anglada-Escudé et al. 2016; Dittmann et al. 2017; Gillon et al. 2017; Bonfils et al. 2018) – all orbit M stars. *Proxima Centauri*, a cool M6V dwarf only 1.3 parsecs from the Sun, harbours a planet in its HZ with a minimum mass of 1.3 Earth masses that receives about 65 per cent of Earth's solar flux (Anglada-Escudé et al. 2016). At 3.4 parsecs from the Sun, the planet *Ross 128b*, with a minimum mass of about 1.4 Earth masses, orbits in the HZ of its cool M4V dwarf star (Bonfils et al. 2018). The *TRAPPIST-1* planetary system of seven transiting Earth-sized planets around a cool M8V dwarf star, which has several (three to four) Earth-sized planets in its HZ, is only about 12 parsecs from the Sun (e.g. Gillon et al. 2017; O'Malley-James & Kaltenegger 2017;

★ E-mail: jomalleyjames@astro.cornell.edu

Ramirez & Kaltenegger 2017). The planet *LHS 1140b* orbits in the HZ of its cool M4V dwarf star, with a measured rocky composition based on its radius of 1.4 Earth radii and mass of 6.7 Earth masses (Dittmann et al. 2017). These four planetary systems already provide several interesting close-by potentially habitable worlds for further study.

The atmospheres of *TRAPPIST-1-d*, *-e*, *-f*, and *-g* have already been observed using Hubble (de Wit et al. 2018), allowing cloud-free hydrogen dominated atmospheres to be ruled out for all but *TRAPPIST-1-g*. More detailed characterization of the atmospheres of these, and other transiting M star HZ planets will be possible with the next generation of observatories, specifically in the visible by telescopes like the extremely large telescopes (ELTs) (see e.g. Snellen et al. 2013; de Wit et al. 2018).

*Proxima Centauri* is an active M6V star, *Ross 128* is an inactive M4V star, *TRAPPIST-1* is a moderately active M8V star and *LHS 1140* is an M4V star with (presently) unknown activity levels. Planets in M star systems can face potential barriers to habitability as a result of their host star's activity (see e.g. discussion in Scalo et al. 2007; Tarter et al. 2007; Shields, Ballard, & Johnson 2016; Kaltenegger 2017). Planets that receive high doses of UV radiation are generally considered to be less promising candidates in the search for life (see e.g. Buccino, Lemarchand, & Mauas 2006). However, several teams have made the case that planets in the HZ of M stars can remain habitable, despite periodic high UV fluxes (see e.g. Heath et al. 1999; Buccino, Lemarchand, & Mauas 2007; Scalo et al. 2007; Tarter et al. 2007; Rugheimer et al. 2015b; O'Malley-James & Kaltenegger 2017).

Two recent studies (Rugheimer et al. 2015a, b) model the amount of radiation reaching the surface of an Earth-like planet with a 1 bar surface pressure through its geological evolution (following Kaltenegger, Traub, & Jucks 2007) for different star types. The study also assessed the biological impact of that radiation for atmospheres corresponding to different times in the geological history of a planet, modelled on Earth.

Generally, an Earth-like planet orbiting an inactive M star receives a lower UV flux than Earth (Rugheimer et al, 2015a, b). However, around active M stars, such planets would be subject to periodic bursts of UV radiation, because of the proximity of the HZ to the star. This increases the surface UV flux on a HZ planet by up to two orders of magnitude at the flare peak for the most active M stars (Segura et al., 2010; Tilley et al. 2019). M stars also remain active for longer periods of time compared to the Sun (see West et al., 2011).

Note that currently we cannot model the expected surface pressure on an exoplanet. A decrease in surface pressure or atmosphere mass increases the UV flux reaching the surface, assuming the same atmospheric composition of a planet (see O'Malley-James & Kaltenegger 2017). Fig. 2 shows the stellar input spectra for the modelled stars, normalized to the incident flux at Earth's equivalent orbital distance as well as model UV fluxes for active M0 to M8 star models (see Rugheimer et al. 2015b). The UV activity in the M6 star model is based on observations of *Proxima Centauri*.

The close proximity of planets to their host star in the HZs of cool stars can cause planetary magnetic fields to be compressed by stellar magnetic pressure, reducing a planet's ability to resist atmospheric erosion by the stellar wind (e.g. Vidotto et al. 2013). A reduction in atmospheric density could result in higher fluxes of UV radiation reaching the planet's surface (e.g. O'Malley-James & Kaltenegger 2019), although erosion could also facilitate an increase in atmospheric oxygen and subsequently ozone, which might reduce surface UV levels. However, planets in the HZs of M stars are also subject to stellar particle fluxes orders of magnitude stronger than those in the solar HZ (Cohen et al. 2014) that could erode their protective ozone shield as well as some of the atmosphere (see e.g. Segura et al. 2010; Bourrier et al. 2017; Wheatley et al. 2017; Tilley et al. 2019).

When UV radiation is absorbed by biological molecules, especially nucleic acids, harmful effects, such as mutation or inactivation can result, with shorter UV wavelengths having the most damaging effects (see e.g. Voet et al. 1963; Diffey 1991; Matsunaga, Hieda, & Nikaido 1991; Tevini 1993; Cockell 1998; Kerwin & Remmele 2007). On the present-day Earth, the ozone layer prevents the most damaging UV wavelengths, UV-C radiation (100–290 nm), from reaching the surface. However, on other HZ planets, a protective ozone layer may not be present; the early Earth, for example, lacked a significant ozone layer (see e.g. Zahnle et al. 2007).

Yet, even in the absence of an ozone layer, depending on the atmospheric composition of a planet, other atmospheric gases, such as sulfur compounds, $CO_2$, or photochemical hazes can absorb UV radiation (see e.g. Cockell et al. 2000; Rugheimer et al. 2015a; Arney et al. 2016; O'Malley-James & Kaltenegger 2017). However, the thinner the atmosphere of a planet is, the higher the damaging UV flux reaching the surface is. Hence, mechanisms that protect biota from such radiation are a crucial part of maintaining surface habitability, especially on planets with thin atmospheres, e.g. less massive planets or moons that cannot maintain a dense protective atmosphere (O'Malley-James & Kaltenegger, 2017, 2019).

On Earth, biological mechanisms such as protective pigments and DNA repair pathways can prevent, or mitigate, radiation damage (see e.g. Cockell 1998; Heath 1999; Neale & Thomas 2016). Additionally, subsurface environments can reduce the intensity of radiation reaching an organism (see e.g. Heath 1999; Wynn-Williams et al., 2002; Ranjan & Sasselov 2016). Some teams have suggested that life that is constrained to habitats underwater, or beneath a planet's surface, may not be detectable remotely (see e.g. Cockell 2014), making an inhabited planet appear uninhabited. However, photoprotective biofluorescence, in which protective proteins absorb harmful UV wavelengths and re-emit them at longer, safer wavelengths, is a mechanism that could enhance detectability (O'Malley-James & Kaltenegger, 2018b). This could result in a temporal biosignature on planets orbiting active M stars. On Earth, observations of corals subjected to intense light regimes suggest that some coral species use such a mechanism to reduce the risk of damage to symbiotic algae, which provide the coral with energy (see e.g. Salih et al. 2000; Roth et al. 2010): fluorescent pigments and proteins in corals absorb blue and UV photons and re-emit them at longer wavelengths (see Section 1.1).

Here, we extend the concept of a fluorescent biosignature originally described in O'Malley-James & Kaltenegger (2018b) to encompass the idea of a new temporal biosignature for planets orbiting active M stars. We outline the possibility of detecting a temporal fluorescent biosphere responding to stellar UV flare events in the HZs of active M stars. In an M star system, such additional emitted visible light would increase the planet's visible flux temporarily at peak emission wavelengths. Especially for a planet orbiting the coolest M stars, which do not emit much flux at visible wavelengths (see Fig. 2), an emitted visible biofluorescent flux could increase the absolute visible flux from a planet substantially, because this additional flux is coupled to the UV flux of the star, not its visible flux. We also explore whether a standard astronomy tool to characterize stellar objects, a colour–colour diagram, can be used to distinguish planets with and without biofluorescent biosignatures and prioritize planets for follow-up spectral observations (following Hegde & Kaltenegger 2013).

### 1.1 Biofluorescence on earth

On Earth, fluorescence – the emission of light by a substance that has absorbed light of a shorter wavelength – is widespread in the natural world. Biofluorescence can be observed in organisms containing biomolecules that fluoresce, such as chlorophyll in vegetation, green fluorescent proteins (GFPs) and GFP-like proteins in a variety of marine life, or fluorescent compounds/pigments in some land animals, such as insects and amphibians (Sparks et al. 2014; Gruber & Sparks 2015; Gruber et al. 2015; Holovachov 2015; Middleton et al. 2015). Note that biofluorecsence differs from bioluminescence, which involves exploiting chemical reactions to generate light and is independent of the radiation environment an organism is exposed to.

A fluorescent object will fluoresce as long as there is a source of radiation that excites an orbital electron in fluorescent proteins, or pigments to its first excited singlet state. Fluorescence emission occurs when this electron relaxes to its ground state. Some of the absorbed energy is released as heat, and some via the emission of a longer wavelength (lower energy) photon than the photon that was originally absorbed; a process known as the Stokes shift. Fluorescence is near-instantaneous (taking place over nanosecond time-scales); however, a related process, phosphorescence involves a different relaxation pathway that results in the delayed emission of light over the course of minutes or hours. The strength of the emitted flux depends on the light environment in the absorption wavelength range and the efficiency of the fluorescence process (or quantum yield), defined as the number of photons emitted to the number absorbed. An efficiency of 1.0 means that each photon absorbed results in an emitted photon, achieving the strongest possible fluorescent flux for a given substance.

On Earth, Biofluorescence in vegetation produces a planet-wide effect that can be detected from orbit (Joiner et al., 2011; Wolanin et al., 2015). This detectable biofluorescence signal on Earth (Joiner et al., 2011) is caused by the fluorescence of chlorophyll in vegetation and can account for up to 1–2 per cent of the strength of the vegetation reflection signal at the fluorescence emission peak wavelength for chlrophyll a, due to vegetation's extensive coverage of the land surface.

As summarized in O'Malley-James & Kaltenegger (2018b) biofluorescence on Earth is also observed in corals, which can fluoresce with a higher fluorescence efficiency than vegetation. Precise, high-resolution observations from Earth-orbiting satellites can disentangle such small signals; however, compared to atmospheric biosignatures or other surface features (see review by Kaltenegger 2017), signals that change the overall planetary flux by less than a per cent will not readily be observable for Earth-like planets orbiting other stars. Yet, given the range of exoplanet radiation environments and potential evolutionary histories, biofluorescence could be a potentially detectable spectral surface biosignature. In a high-UV environment, such biofluorescent life could evolve to become widespread and to fluoresce even more efficiently, producing a remotely detectable signal.

For our models we concentrate on the well-studied biofluorescence in corals. Scleractinian (hard) coral reefs are comprised of colonies of identical animals called polyps (small sac-like animals with a set of tentacles surrounding a central mouth) that secrete calcium carbonate to form a hard exoskeleton. Excessive light can be harmful to corals, either by damaging the algal photosystem, or by increasing oxidative stress, via photochemical reactions (Bhagooli & Hidaka 2004; Takahashi & Murata 2008; Roth et al. 2010). Some species contain fluorescent pigments and proteins with excitation spectra in the UV-A (315–400 nm) and blue regions of the spectrum, which have emission maxima between 420 and 700 nm.

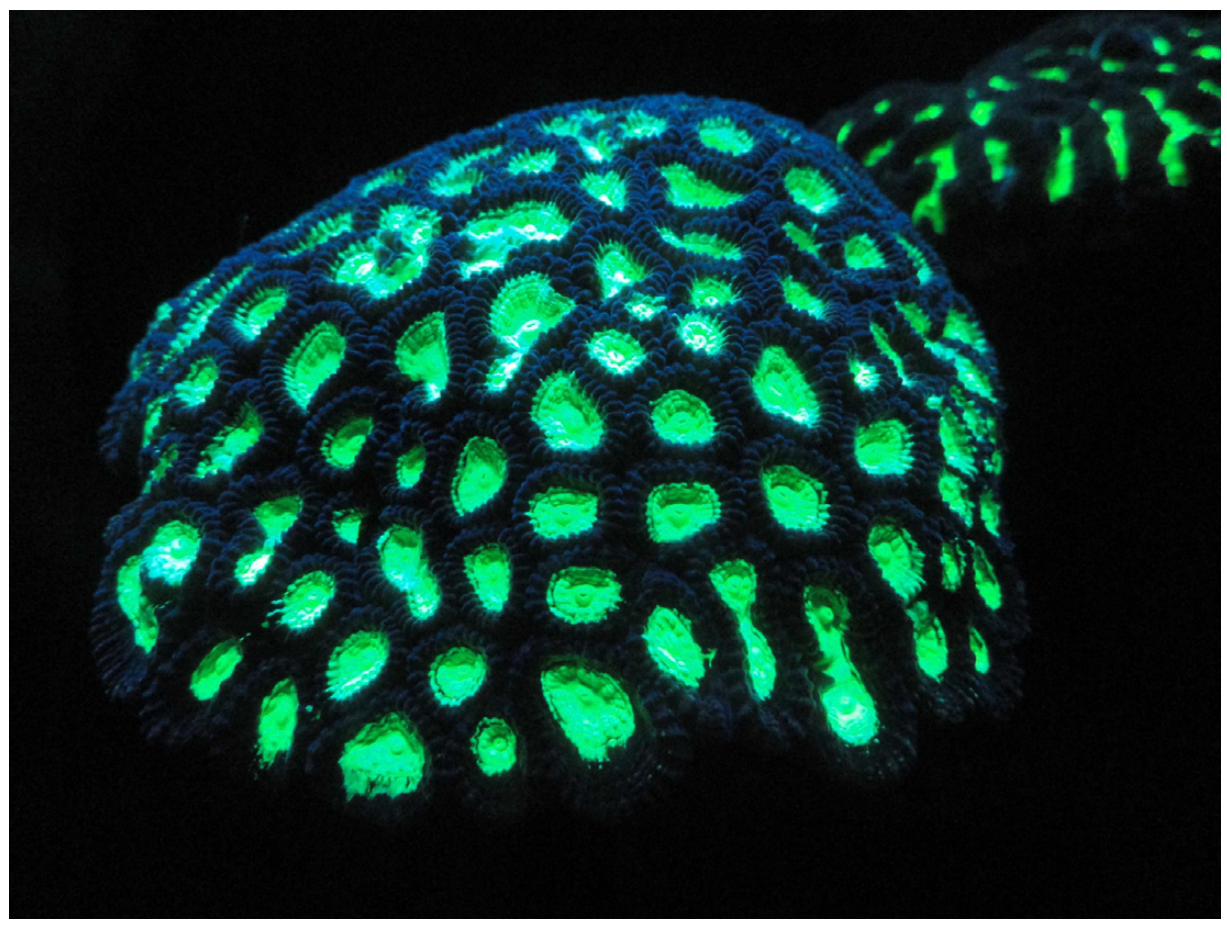

**Figure 1.** An example of coral fluorescence. Coral fluorescent proteins absorb near-UV and blue light and re-emit it at longer wavelengths (see e.g. Mazel & Fuchs 2003). *Image made available under Creative Commons CC0 1.0 Universal Public Domain Dedication.*

UV absorption by these pigments/proteins is a possible protection mechanism for symbiotic algae from harmful UV radiation, see Figs 1(b) and 2 (see e.g. Salih et al. 2000; Gorbunov et al. 2001).

The visible signature of a coral reef has two components: elastic scatter (reflected light) and inelastic scatter (fluorescent light). Fluorescence can be a significant factor in the appearance of coral reefs (Fuchs 2001). The magnitude of the increase in intensity at the emitting wavelengths of fluorescent pigments and proteins varies depending on the intensity of the radiation absorbed, the wavelengths a pigment or protein absorbs/emits, and the spectral overlaps between pigments and proteins with different excitation spectra within a reef (Fuchs 2001). The four most commonly occurring fluorescent pigments/proteins in corals have approximate fluorescence emission peaks that cover the full range of the visible spectrum at 486 (cyan), 515 (green), 575 (orange), and 685 nm (red). The cyan and green fluorescence is due to GFP-like proteins, the orange fluorescence is caused either by a GFP-like protein or a phycoerythrin from symbiotic cyanobacteria, while the red fluorescence is due to chlorophyll in symbiotic algae (Zawada & Mazel, 2014). The properties of these pigments/proteins are summarized in Table 1. The absorption ranges for the fluorescent pigments and proteins used in this study are superimposed over the fluxes plotted in Fig. 2(b).

The fluorescent efficiencies of coral fluorescent pigments and proteins vary depending on the type of protein and the light environment a species lives in. GFPs are the most efficient forms of fluorescent proteins, with some reaching quantum yields of up to 0.79 (Lagorio, Cordon, & Iriel 2015). GFPs, extracted from jellyfish were the first fluorescent proteins to be studied (Johnson et al., 1962; Shimomura, Johnson, & Saiga 1962; Morin & Hastings 1971; Morise et al. 1974; Tsien 1998). Over time these have been adapted and engineered for use in a variety of applications, from fluorescent microscopy to transgenic pets (see e.g. Stewart 2006 and references therein).

Biofluorescent corals on Earth cover only ~0.2 per cent of the ocean floor, which results in a change in visible flux at a given fluorescent wavelength of a fraction of a per cent in Earth's overall spectrum (see Table 2). This has a negligible impact on the globally averaged spectrum of Earth seen as an exoplanet because of the small surface coverage. However, coral (or similar) biofluorescence could be widespread on other planets. In addition biofluorescence becomes more efficient under artificial lighting conditions (see e.g. Mazel & Fuchs 2003; Hochberg et al. 2004), causing increases in the strength of the biofluorescence signal of up to two orders of magnitude (Table 2); the increase depending on the spectral distribution of the illuminating light source (Mazel & Fuchs 2003). For example, laser-induced fluorescence enables corals to be successfully identified and monitored from the ocean surface, rather than in situ (Myers et al. 1999; Mumby et al. 2004). Hence, more intense UV environments, like those found on planets orbiting active M stars, should result in a higher fluorescence efficiency and thus a stronger fluorescence effect.

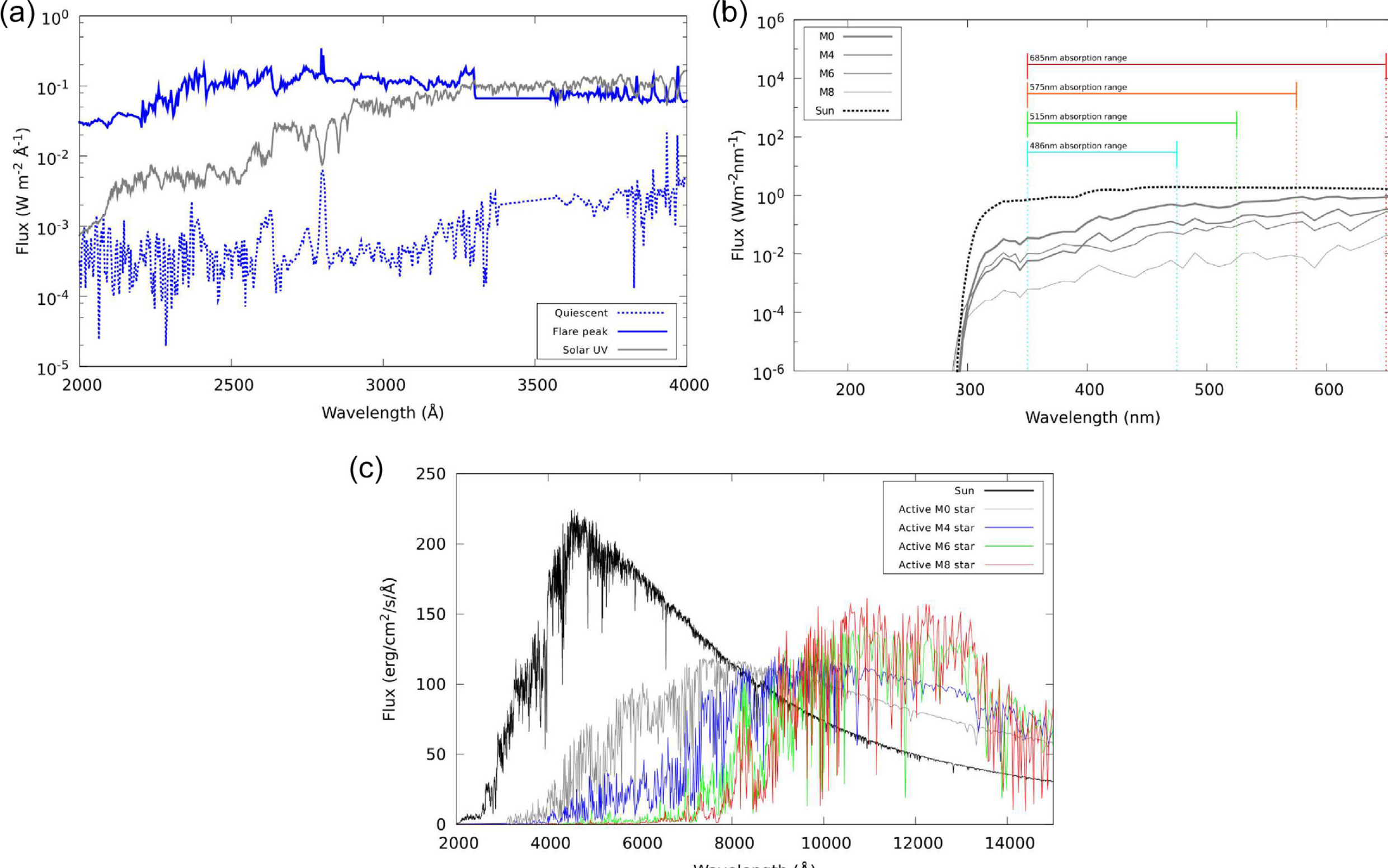

**Figure 2.** (a) The UV flux at the top of the atmosphere of a planet at a 1 au equivalent distance orbiting the active M star AD Leo during quiescence (blue dashed) and flaring (blue solid) (Segura et al. 2010, using flare observations from Hawley & Pettersen 1991). The solar UV spectrum is shown for comparison (grey). (b) Model UV surface fluxes for Earth analogue planets orbiting active M0 to M8 stars at a 1 au equivalent distance, compared to the solar flux at Earth's surface. The M0 and M4 flux models are scaled to the activity of AD Leo (see Rugheimer et al. (2015b). The UV activity in the M6 star model is based on observations of *Proxima Centauri*. The absorption ranges for the fluorescent proteins used in this study are superimposed over the fluxes. (c) Stellar input spectra for the modelled stars, normalized to the incident flux at Earth's orbit.

### 1.2 UV increases during M star flares

Our proposed temporal biofluorescent signature around an active M star is induced by UV flares and the corresponding UV flux levels on the surface of a planet. Corals fluoresce brightly under a range of light regimes on Earth. UV-A wavelengths are the main UV excitation wavelengths for coral fluorescence. Many fluorescent corals are found at shallow ocean depths where the UV-A regime is similar to the surface flux on Earth. However, bright coral fluorescence is observed down to depths of 50–60 m (Eyal et al. 2015), where the UV-A flux can be as low as 12 W m$^{-2}$ (Eyal et al. 2015). The modelled UV flux reaching the surface of a planet analogue to the present-day Earth, orbiting a quiescent M star at a 1 au-equivalent distance varies between 0.04 and 1.36 W m$^{-2}$, depending on spectral type for a 1 bar surface pressure (Rugheimer et al. 2015a). This is 1–2 orders of magnitude below the modelled UV flux on present-day Earth's surface (32.3 W m$^{-2}$; Rugheimer et al. 2015a), respectively.

**Table 1.** The four most common fluorescent pigments and proteins in coral species. The quoted fluorescent efficiencies are from coral species selected for being highly fluorescent (Mazel & Fuchs 2003; Mazel et al. 2003).

| Emission peak (nm) | Excitation range (nm) |
|---|---|
| 486 | 350–475 |
| 515 | 350–525 |
| 575 | 350–575 |
| 685 | 350–650 |

For the active M star *AD Leo*, the modelled quiescent UV surface flux on an Earth analogue planet with 1 bar surface pressure in the HZ is 2.97 UV-A, 0.01 UV-B, and 2.13 × 10$^{-14}$ W m$^{-2}$ UV-C (Segura et al. 2010), an order of magnitude lower than on present-day Earth (consistent with results by Rugheimer et al. 2015b). However at the flare peak, UV-A and UV-B levels reach 120.77 UV-A and 3.15 W m$^{-2}$ UV-B (Segura et al. 2010), an order of magnitude higher than on present-day Earth.

Models of the UV irradiation reaching the surface of an Earth-like planet with a 1 bar surface pressure (see Rugheimer et al. 2015a, b for details) through its geological evolution (following Kaltenegger et al. 2007) show that for a younger Earth the UV surface flux increases. Lower surface pressure also increases the UV levels on a planet's surface for atmospheres with similar composition (O'Malley-James & Kaltenegger, 2017, 2019).

**Table 2.** The percentage change in reflected + emitted flux from a planet's surface at a 1 au equivalent distance before and during fluorescence at each of the peak emission wavelengths of the four common marine fluorescent pigments/proteins in Table 1 for M0, M4, M6, and M8 type M stars (equivalent to the hottest M star (M0), *LHS-1140, Proxima Centauri*, and *TRAPPIST-1*, respectively). We show a range of surface coverage scenarios. The values shown are for 100 per cent efficient fluorescence (except for the Earth analog scenario), but can be directly scaled to quantify lower efficiency cases. Note that we do not include the planet orbiting *Ross-128* here, because it is currently classified as an inactive star. ∗Based on common fluorescent pigment/protein efficiencies and assuming no water attenuation.

| Host star | Planetary flux increase (in per cent) | | | | | |
|---|---|---|---|---|---|---|
| | No clouds: 100 per cent biofl. surface at wavelength (nm) | | 50 per cent clouds: 100 per cent biofl. surface | No clouds: 70 per cent ocean 30 per cent biofl. surface | 50 per cent clouds: 70 per cent ocean 30 per cent biofl. surface | Earth analogue: 50 per cent clouds 70 per cent ocean 29.8 per cent surface 0.2 per cent biofl.[a] |
| M0 (example of the hottest M stars) | 486 | 1340 | 260 | 700 | 150 | 0.050 |
| | 515 | 1350 | 270 | 770 | 160 | 0.100 |
| | 575 | 700 | 260 | 550 | 160 | 0.100 |
| | 685 | 800 | 240 | 600 | 150 | 0.020 |
| M4 (LHS-1140) | 486 | 1100 | 210 | 570 | 123 | 0.041 |
| | 515 | 1100 | 220 | 630 | 131 | 0.082 |
| | 575 | 570 | 210 | 450 | 131 | 0.082 |
| | 685 | 660 | 200 | 500 | 123 | 0.016 |
| M6 (Proxima Centauri) | 486 | 1030 | 200 | 540 | 115 | 0.039 |
| | 515 | 1040 | 210 | 600 | 123 | 0.077 |
| | 575 | 540 | 200 | 420 | 123 | 0.077 |
| | 685 | 620 | 180 | 460 | 115 | 0.015 |
| M8 (TRAPPIST-1) | 486 | 857 | 166 | 450 | 95 | 0.032 |
| | 515 | 864 | 170 | 490 | 100 | 0.064 |
| | 575 | 250 | 166 | 350 | 100 | 0.064 |
| | 685 | 510 | 150 | 380 | 95 | 0.013 |

[a]Based on common fluorescent pigment/protein efficiencies and assuming no water attenuation.

The UV surface values would increase when one accounts for the deformation of a planet's compressed magnetic field due to stellar magnetic pressure (Lammer et al. 2007), which would increase the risk of atmospheric erosion and subsequent atmospheric mass-loss. Increased UV flux on a planet's surface could also occur due to associated proton events with the flare, in which charged particles from the star are accelerated in the direction of a flare. If the planet is in the line of sight of the charged particle stream, ionized particles can interact with the planet's atmosphere, forming odd hydrogen species and breaking up $N_2$ to form $NO_x$ species, both of which can destroy ozone (see e.g. Segura et al. 2010; Grenfell et al. 2012). Accounting for a proton event associated with an single M star flare, Segura et al. (2010) calculated that approximately 70 d after the flare, ozone depletion could cause surface UV fluxes to increase for up to 2 yr. If multiple flare events are accounted for, an Earth-like atmosphere could be almost completely depleted of ozone within 10 yr (Tilley et al. 2019). Frequent flares and proton events could thus permanently weaken, or erode, a planet's ozone layer and overall atmosphere, allowing more UV radiation to reach the planet's surface if the star's activity cycle is shorter than the recovery time of the atmosphere.

## 2 METHODS

We combine surface spectra for a fluorescent coral biosphere and Earth's ocean to generate a reflectance spectrum for our model planets. We assume the biofluorescent surface biosphere is living in a shallow, transparent ocean. This is consistent with shallow-water coral reef habitats on Earth, which are typically warm, clear seawater environments, at depths as shallow as 0–3 m (Kleypas, McManus, & Meez 1999; Nagelkerken et al. 2000). Furthermore, the lower flux of Earth-like photosynthetically active radiation (PAR) on an M star planet would make shallow water habitats more preferable for any aquatic photosynthesizers. We can adjust the ratio of coral-to-ocean spectra in our model surface to simulate different fractions of uninhabited deep ocean versus inhabited shallow waters.

Using fluorescent protein absorption and emission models, we can turn fluorescence 'on' and 'off' in our model surface to show the change in apparent reflectance (the sum of reflected and emitted light) this causes. We assume that evolution under a high, fluctuating UV radiation environment will drive surface life to develop efficient fluorescence. Therefore we model biofluorescence that has an efficiency (the ratio of absorbed to emitted photons) of 100 per cent, i.e. all incident photons in the excitation range of the protein are absorbed and re-emitted (see Section 2.2).

We then include the effects caused by an Earth-like atmosphere (as it would appear when exposed to an M star stellar flux at a 1 au-equivalent distance) with varying cloud-fractions between 0 and 50 per cent (i.e. from clear skies to an Earth-like cloud coverage) to generate the model spectrum and colour for a set of model planets. We then replace the coral surface with fluorescent mineral surfaces to compare the biofluorescence spectrum with potential false-positive mineral fluorescent signatures. We also compare our biofluorescence model with a model that substitutes the biofluorescent surface biosphere with vegetation to compare their spectra. Full details of our models are provided in the subsections below.

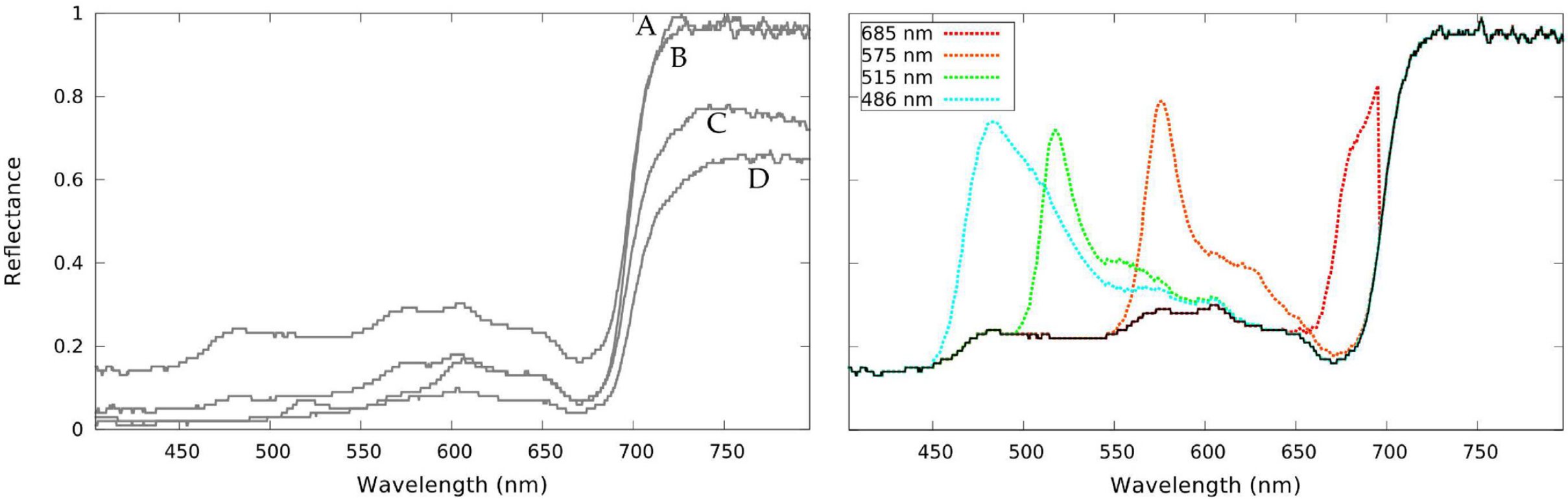

**Figure 3. (Left)** Four sample coral species (A, B, C, and D) spanning the colour distribution of corals that we model as biofluorescent surfaces. **(Right)** An example of modelled fluorescence using coral B and spectral data on fluorescence emission of four common coral fluorescent pigments/proteins (Coral data from Roelfsema. & Phinn 2006; Clark 2007).

### 2.1 Planetary models

The surface UV flux on a planet will determine the strength of a biofluorescence signal from biota on the surface. We use model stellar spectra with UV fluxes scaled to the AD Leo flare peak (Fig. 2) as approximations to the maximum UV output from active M stars of different types – M0, M4, M6, M8 – as top of atmosphere input spectra. Using an Earth-composition model atmosphere, we model transport of radiation through such an atmosphere to estimate the expected surface radiation flux in each case. Our UV surface flux estimates then provide the input for the biofluorescence models described in Section 2.2. We perform these simulations using *EXO-Prime* (see e.g. Kaltenegger & Sasselov 2009); a coupled 1D radiative-convective atmosphere code developed for rocky exoplanets. The code is based on iterations of a 1D climate model (Kasting & Ackerman 1986; Pavlov et al. 2000; Haqq-Misra et al. 2008), a 1D photochemistry model (Pavlov & Kasting 2002; Segura et al., 2005, 2007), and a radiative transfer model (Traub & Stier 1976; Kaltenegger & Traub 2009) that calculates the model spectrum of a rocky exoplanet in the HZ. *EXO-Prime* models exoplanet atmospheres and environments depending on the stellar and planetary conditions, including the UV radiation that reaches the surface and the planet's reflection, emission and transmission spectrum (see Rugheimer et al. 2015a, b for details on surface UV calculations).

We can vary the fractions of different surface coverings (land, ocean, and biofluorescent surface life). Initially, we model planetary surface spectra assuming the surface is a global ocean inhabited by biofluorescent life, or a hypothetical fully vegetation covered planet. This constitutes the ideal case with the strongest signal from a surface biosignature. We then reduce the surface fraction covered by biofluorescent life to test how the spectra and colours change by adding ocean (from the USGS Spectral Library) as an additional surface to explore the effect of different fractions of inhabited versus uninhabited surface on the spectrum. Then we add clouds (following Kaltenegger et al. 2007) to the model. We assume clouds block our view of any surface feature. We model different cloud coverages, up to 50 per cent (an Earth-like cloud fraction). Any surface feature signature is reduced with increasing cloud cover because clouds are highly reflective and therefore strongly influence a planet's spectrum (see e.g. Kaltenegger et al. 2007). We compare the spectra of biofluorescent surfaces to that of the same surface fraction covered in vegetation to compare the difficulty of detection of these two surface biosignatures.

We use *EXO-Prime* to calculate the UV-visible photon surface fluxes for planets orbiting active M0, M4, M6, and M8 type stars at a 1 au-equivalent distance. These are equivalent to the hottest M star type (M0), and the planet-hosting M stars *LHS-1140* (M4), *Proxima Centauri* (M6), and *TRAPPIST-1* (M8). The angular separations of the known HZ planets in these close-by M star systems make them prime targets for the detection of biosignatures, including a temporal fluorescent biosignature, for upcoming ground-based telescopes like the ELT (see Section 4). Using the surface reflectance profiles for our model planets, combined with the absorption/emission behaviour of the fluorescent pigments and proteins (Table 1), we quantify the change in outgoing photon flux at the peak emission wavelengths caused by each of the four common coral fluorescent pigments/proteins during fluorescence. By comparing the spectra (before and during fluorescence) with the expected visible fluxes from the host stars, we can estimate the improvement of the star–planet contrast ratio (based on Traub & Oppenheimer 2010) induced by fluorescence. Note that we do not include *Ross-128* as an example in Table 2 because it is currently classified as an inactive M4 star (Bonfils et al. 2018).

### 2.2 Biofluorescence models

We use the efficiency limits of terrestrial fluorescent proteins as a guide to the exploration of the magnitude of our modelled biofluorescence. GFPs on Earth have been observed in nature to have fluorescent efficiencies of up to 79 per cent (Lagorio et al. 2015). Also, GFPs have been engineered in the lab to have much higher fluorescence efficiencies by taking advantage of useful mutations. This has resulted in proteins with efficiencies of up to 100 per cent (Ilagan et al. 2010; Goedhart et al. 2012). Furthermore, highly efficient fluorescence has been observed in nature, for example, the trees *Pterocarpus indicus* and *Eysenhardtia polystachya* produce matlaline, which fluoresces blue with an efficiency of almost 100 per cent (Lagorio et al. 2015). Therefore we model biofluorescence that has an efficiency (the ratio of absorbed to emitted photons) of 100 per cent, i.e. all incident photons in the excitation range of the protein are absorbed and re-emitted, following the emission profiles in Fig. 3.

We simulated coral fluorescence by using models of the fluorescence response of the four common fluorescent pigments and proteins in corals (shown in Table 1), which have fluorescence peaks at 486, 515, 575, and 685 nm. The excitation range of the common coral fluorescent proteins on Earth does not extend to wavelengths shortward of UV-A wavelengths; hence our models, which are derived from observations of coral fluorescence, are limited to UV-A absorption. However, fluorescent proteins with absorption spectra that encompass higher energy UV-B and UV-C wavelengths would confer the best protection from M star UV flares; we speculate about the possibility of this in the Section 4. Therefore, we compare two fluorescent cases: a bright fluorescence case that assumes the evolution of a mechanism to absorb across the full near-UV spectrum, and a case limited to the excitation range of Earth-like fluorescent proteins, calculating the emission flux based on the stellar photon flux received by a planet at a 1 au-equivalent distance during a flare (see Section 2.1).

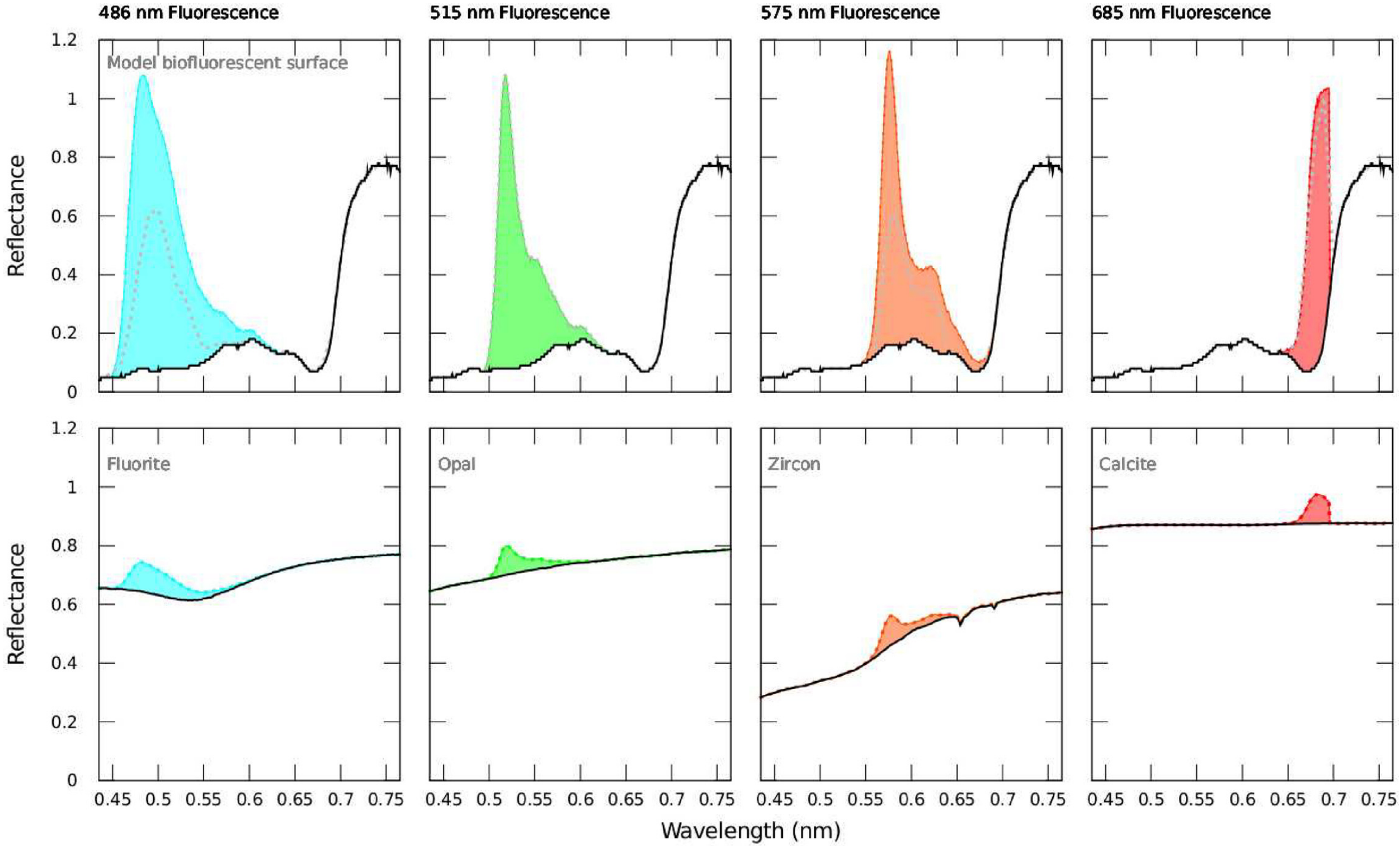

**Figure 4.** Reflectance spectra in the visible for coral, using coral C as an example (top), and fluorescent minerals (bottom). Fluorescence at each of the common coral fluorescent pigment/protein emission wavelengths was simulated for 100 per cent efficiency with absorption profiles limited to the ranges observed on Earth (Table 1) (grey dashed). Note that the apparent reflectance (the sum of reflected and UV-induced-emitted flux) can exceed 1 in some cases. Examples shown are for an M0 star radiation environment. Green fluorescence produces the strongest increase in flux under an active M star surface radiation regime, so we scale the fluorescence responses at other wavelengths to the magnitude of this response (shaded curves), representing a scenario in which the UV absorption range of the fluorescent proteins increases in response to selection pressures imposed by a high-UV environment. The four mineral species shown were chosen for their abilities to fluoresce at similar wavelengths to corals and represent the strongest fluorescent minerals. [Non-fluorescent coral spectra from Roelfsema & Phinn (2006). Mineral spectra sources: *Sources: USGS Digital Spectral Library (Clark 2007), ASTER spectral library, California Institute of Technology. Fluorescence was simulated using data from C. Mazel.]*

We add this simulated fluorescence on to to the reflected flux from our model planet surface. Note that our non-excited coral surface spectra are natural coral spectra that will contain very low fluorescence excitation and emission features caused by Earth's UV radiation environment. However, compared to the UV induced fluorescence signals we model here, the strengths of these features are negligible; hence we add our modelled fluorescence on to the natural spectra, treating them as if they contain no fluorescence features. When fluorescence is added to our model surface spectrum using Earth-like absorption ranges for each of the fluorescent proteins, green fluorescence shows the largest increase (of ∼1000 per cent at the peak emission wavelength) in the visible spectrum (Fig. 4), whereas blue and orange fluorescence show the smallest increase of ∼500 per cent (Fig. 4). To compare fluorescence across the different fluorescent protein colours, we scale each colour to the green fluorescence response. This also accounts for the evolution of a broader UV absorption range for fluorescent proteins in high-UV environments. Note that this is additional emitted light in the visible and can surpass the reflectance value that 100 per cent reflectivity alone would provide.

### 2.3 Colour–colour diagrams as a diagnostic tool

We use a standard astronomy tool to characterize stellar objects, a colour–colour diagram, to explore if one can distinguish planets with and without biofluorescent biosignatures (following Hegde & Kaltenegger 2013). To determine the difference between the visible flux, $r$, of two different colour bands, we use equation (1):

$$C_{AB} = A - B = -2.5\log_{10}(r_A/r_B) \tag{1}$$

where $C_{AB}$ is the difference between two arbitrary colour bands, $A$ and $B$.

We use standard Johnson–Cousins *BVI* broad-band filters to define the colour bands ($0.4\ \mu\mathrm{m} < B < 0.5\ \mu\mathrm{m}$; $0.5\ \mu\mathrm{m} < V < 0.7\ \mu\mathrm{m}$; $0.7\ \mu\mathrm{m} < I < 0.9\ \mu\mathrm{m}$). We used colour–colour plots to explore colour change on model planets caused by flare-induced fluorescence compared to vegetation for clear atmospheres, cloudy conditions, and different ocean fractions.

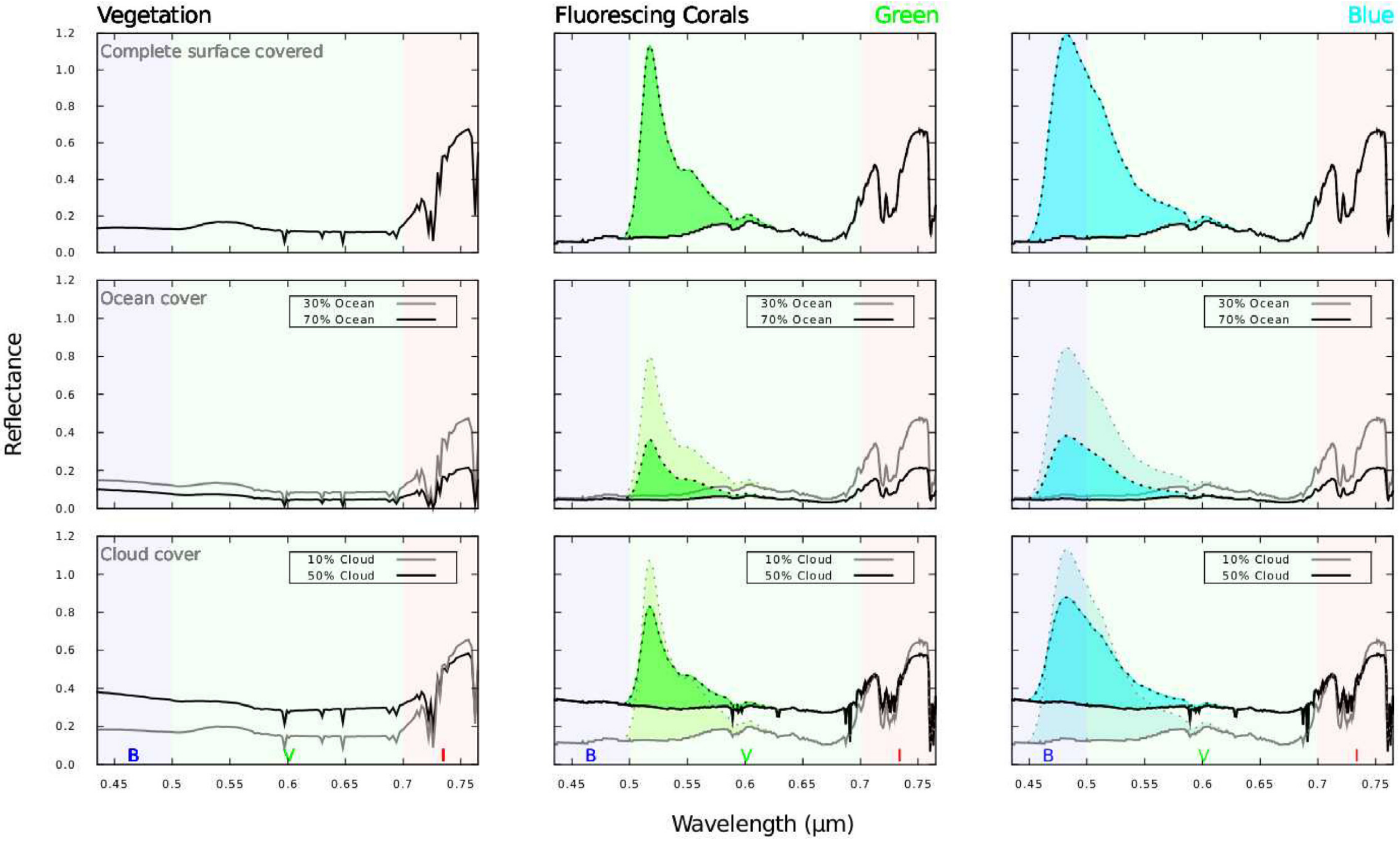

**Figure 5.** Comparison of spectra for a planet with vegetation (left) and biofluorescent (right) surfaces (100 per cent efficient with a broad UV absorption range under an active M0 star radiation environment). A present-day Earth atmosphere has been added to all models shown. We use a coral spectrum (coral C) with modelled fluorescence (515 nm fluorescence is used in this example). (Top) Surface biosignatures are assumed to cover 100 per cent of the planet. (Middle) An ocean fraction of 30 per cent and 70 per cent is added, reducing the surface biosignature fraction to 70 per cent and 30 per cent, respectively. (Bottom) Cloud cover fractions of 10 per cent and 50 per cent are added to the model, assuming the surface of the planet is completely covered with the biosignature (like the top panel) to show the effects of clouds separately from surface fraction coverage (middle).

### 2.4 Investigating false positives

It is also possible for fluorescence to occur abiotically. Some minerals (e.g. calcite, fluorite, opal, zircon) and polycyclic aromatic hydrocarbons (PAHs; e.g. fluoroanthene, perylene, pyrene) are fluorescent at similar wavelengths to those of fluorescent corals; a result of metal cation impurities in the case of minerals, and delocalized electrons in aromatic molecule groups for the case of PAHs (see e.g. McDougall 1952; Modreski 1987; Beltrán, Ferrer, & Guiteras 1998). Therefore, we explore how to distinguish biofluorescence from fluorescent minerals.

Fig. 4 compares spectra from biofluorescent models with spectra from fluorescent mineral surfaces. The four mineral species were chosen for their abilities to fluoresce at similar wavelengths to corals and represent the strongest fluorescent minerals (see e.g. Modreski 1987; fluorescence information from: Luminescent Mineral Data base[1]). The lower row of Fig. 4 shows reflectance spectra for simulated coral fluorescence for four wavelengths compared to minerals that fluoresce at similar wavelengths in the right-hand column for a clear atmosphere. The spectra show that both have different responses to UV flux. The signal of fluorescent minerals has a different shape than the biofluorescent signal and, under the assumption that biofluorescence evolves, the mineral signature is also much weaker. Hydrocarbons and the metal inclusions within fluorescent minerals would not be subject to Darwinian evolution, so we assume here that their fluorescent levels would be comparable to those on Earth or slightly increased, or decreased, scaled to the UV flux at the exoplanet's surface compared to Earth's surface.

[1]http://flomin.org

## 3 RESULTS

When we add fluorescence to our model surface spectrum using Earth-like absorption ranges for each of the fluorescent proteins, green fluorescence shows the largest increase (of ∼1000 per cent at the peak emission wavelength) in the visible spectrum (Fig. 4), whereas blue and orange fluorescence show the smallest increase of ∼500 per cent (Fig. 4). To compare fluorescence across the different fluorescent protein colours, we scale each colour to the green fluorescence response. This also accounts for the evolution of a broader UV absorption range for fluorescent proteins in high-UV environments. Note that this is additional emitted light in the visible and can surpass the reflectance value that 100 per cent reflectivity alone would provide.

### 3.1 Biofluorescence can produce a strong change in the visible spectrum of an Earth-like planet

Biofluorescence can increase the visible flux in the planet's spectrum by approximately over an order of magnitude at peak fluorescence emission wavelengths (see Table 2) compared to a non-fluorescent state for an Earth-like planet with clear skies and a surface that is fully covered by a highly efficient fluorescent biosphere (Fig. 5, top). We show two fluorescent biota as examples of the four most common biofluorescent proteins. For comparison we also show another surface biofeature (Fig. 5, left), vegetation, assuming similar surface coverage of both biospheres. Biofluorescence changes the spectrum of our model planets more strongly than vegetation (Fig. 5).

Table 2 shows the maximum flux that such biofluorescence could cause in order to explore the feasibility of detecting such a signal, compared to detecting reflected light from an Earth-like exoplanet. We show results for known exoplanets in the HZ orbiting three of the closest four M stars [*Proxima-b* (M6), *LHS 1140b* (M4), and the *TRAPPIST-1* (M8) planets]. We did not include *Ross 128b* because its host star is currently not an active star. For comparison we add an M0 host star as well, to compare to planetary models for the hottest M star host. If strong biofluorescence evolves on the surface of these planets, which are exposed to a varying UV environment, it would produce an increase in temporal brightness in the planet's visible spectrum at the peak biofluorescence emission wavelengths for the different fluorescent proteins depending on the SEDs of the host stars that induce the biofluorescence.

We compare the increase of the flux at peak wavelength for models of a widespread biofluorescent biosphere under different scenarios. We first quantify the maximum flux, which would be received for a clear atmosphere and 100 per cent surface coverage (column 1 of Table 2), leading to an increase in flux at the peak fluorescence emission wavelength of between 1300 per cent and 200 per cent. Then we compare that to modelled planets with 50 per cent cloud coverage (the average cloud coverage on Earth) leading to a flux increase of between about 250 per cent and 150 per cent compared to no fluorescence. The third model, given as an example in column 3 of Table 2, is a planet model covered in 70 per cent uninhabited ocean, with the remaining 30 per cent of the surface covered with biofluorescent life (similar to the land: ocean fraction on Earth). In this case the increase in the flux at peak wavelength caused by biofluorescence is between about 350 per cent and 700 per cent, compared to no fluorescence (Fig. 5 middle) for a clear atmosphere and between 150 per cent and 100 per cent for 50 per cent clouds. For reduced surface coverage of biofluorescent biota, the fluorescent signal strength reduces according to the surface fraction in our models, assuming a similar signal per surface area. We have included a model for an Earth-analogue planet with only a 0.2 per cent surface fraction of biofluorescent corals and 50 per cent cloud coverage for comparison to show how surface fraction coverage reduces the signal and why the signal would be very small for Earth seen as an exoplanet; only between 0.01 per cent and 0.1 per cent. As discussed earlier, Earth's biosphere did not evolve in a high UV radiation environment, such as the high UV radiation environments suggested for the surfaces of planets orbiting active M stars. Evolution under these conditions could increase the surface coverage of biota adapted to high-UV fluxes, like a biofluorescent biosphere. The effect of changing cloud cover (Fig. 5 bottom) reduces the effect of any surface reflection features on the overall spectrum, because it would block light from the surface, see Table 2.

We call the effects out separately in Fig. 6 to show their individual behaviour for the four biofluorescent pigments/proteins. The effect of surface fraction (columns) and cloud coverage (inlay) are similar for all surface features. Note that the detectability of any surface feature will depend on the surface fraction coverage as well as the cloud coverage.

### 3.2 A biofluorescent biosphere can be identified in a colour–colour diagram to prioritize follow-up observations

Biofluorescence is separated in traditional astronomical colour–colour space from other surface biosignatures, fluorescing minerals, and other Solar system planet colours. Fig. 4 shows that the visible spectra of biofluorescence and fluorescent minerals differ considerably and can be distinguished with low-resolution spectroscopy of a planet. Fig. 5 shows two biofluorescent models and compares the shape and the change in the visible spectra for surfaces covered by biofluorescent biota and vegetation, as well as the effect of clouds and fractional surface coverage. We calculate colours from these spectra to explore whether a colour–colour diagram can be used to initially identify biofluorescent biospheres and help prioritize extrasolar planets for follow-up (following Hegde & Kaltenegger 2013). Note that we derive these colours from theoretical spectra and thus do not add error bars to our calculations. Our colour–colour diagrams are intended to provide input for instrument simulators for upcoming instruments built to detect the colours of rocky planets, which will have individual errors, depending on the telescope and instrument used.

We used the planet model spectra (Figs 4 and 5) with and without clouds, and with varying surface fractions of biological versus ocean surface, to create standard Johnson colours from their fluorescent and non-fluorescent spectra for biofluorescence as well as for fluorescent minerals. We add Solar system planets [using spectra from Irvine et al. (1968) and Karkoschka (1994) for Mars, Venus, Earth, Jupiter, Saturn, Uranus, and Neptune] for comparison to explore false-positives in the colour space.

In colour–colour space there is no conflict (i.e. false positives) with known Earth-based fluorescent minerals. The distinct colour change of a fluorescent biosphere from a non-fluorescent to a fluorescent state in a colour–colour diagram (Fig. 6) can be used to initially distinguish planets with and without temporal biofluorescent biosignatures. We show the movement in colour space for 50 per cent and 100 per cent fluorescence efficiency. The biofluorescent proteins at 515 and 575 nm show the widest spread in colour space for the fluorescent stage. Adding clouds to the model reduces the amount of colour change compared to the clear-sky cases because clouds block light from the surface (Fig. 6 inset panels). We use a full surface coverage model (Fig. 6, left), e.g. an ocean world with a biofluorescent marine biosphere that covers the whole planet, as well as a biofluorescent-to-uninhabited surface fraction similar to Earth's land: ocean fraction (Fig. 6, right) to show the range of colour space such planets can occupy and discuss false positives. We use a clear sky case and a 50 per cent cloud fraction case (based on Earth's cloud fraction) in our examples to also show the changes clouds introduce in colour space for these models.

In colour–colour space there is no conflict (i.e. false positives) with Solar system planets, except for a 100 per cent fluorescence clear-sky scenarios for the colour of Mars. For full surface coverage and clear-skies the biofluorescent surfaces are distinct from false positives in colour space, before and during fluorescence at the four common fluorescent protein emission wavelengths for all proteins for 50 per cent efficiency (Fig. 6). Only for 100 per cent fluorescent efficiency the biofluorescent proteins at 515 and 575 nm for the fluorescent stage of Coral B, are very close to Mars' colour position, making the change in colour essential to distinguish these two fluorescent stages from a Mars like object for 100 per cent efficiency in fluorescence. Note that using 100 per cent efficiency was the only case where this confusion with the colour of another Solar system objects occurs (for efficiencies from 0 per cent to 100 per cent);

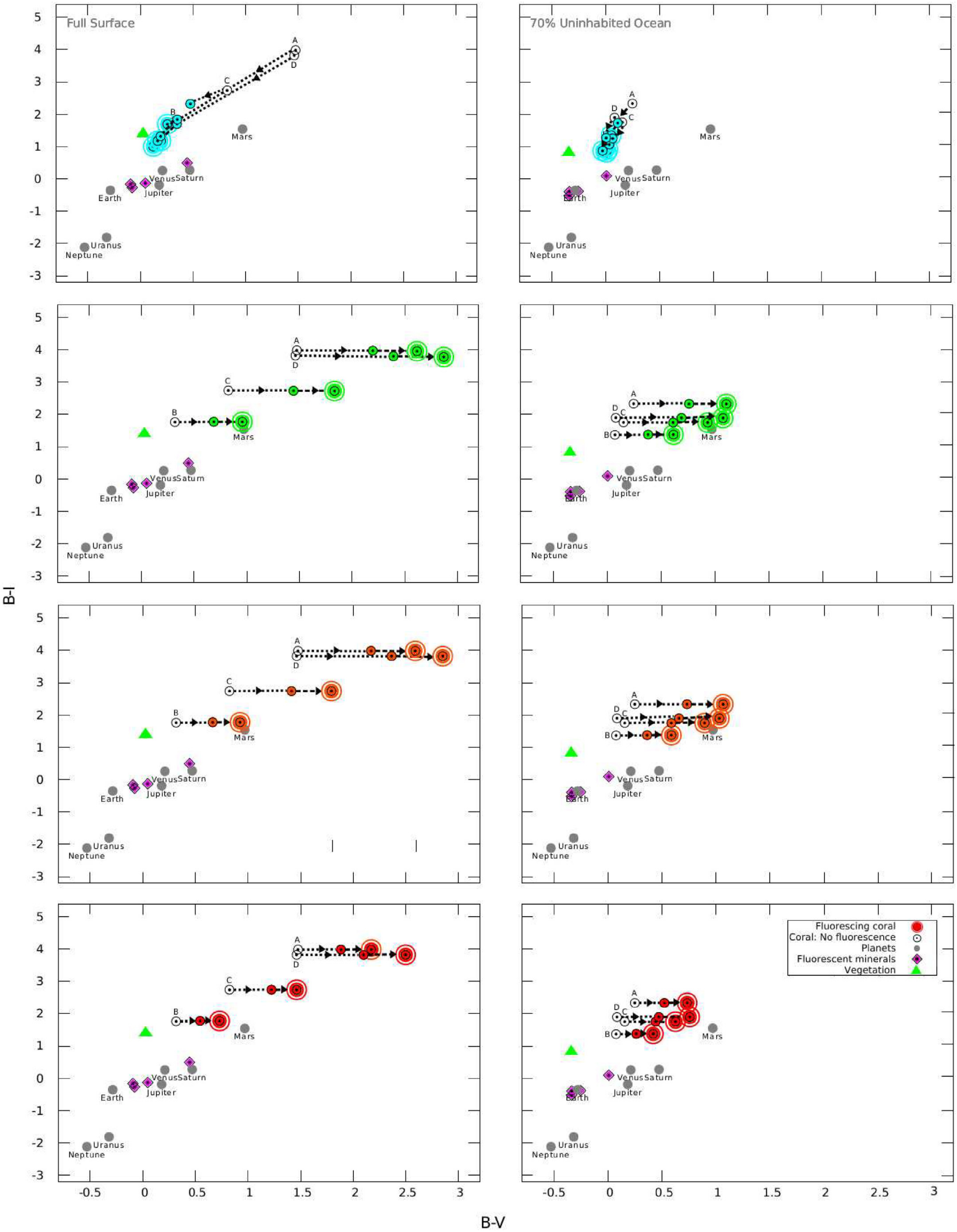

**Figure 6.** Colour–colour diagrams showing the surface colour of our planet models with clear atmospheres under an M0 star radiation environment. (Left column) surfaces that are completely covered by biofluorescent corals. (Right column) surface with 70 per cent uninhabited ocean fraction. Non-fluorescent corals (A, B, C, and D) are indicated by unfilled circles. Fluorescing corals are indicated by the shaded circles [the colour of the shading represents the fluorescent wavelength described in a particular panel: cyan (486), green (515), orange (575), red (685 nm)]. The direction of movement through the colour space as the surface transitions from a non-fluorescent to a fully fluorescent state is indicated by dashed lines. The large shaded circles show the bright 100 per cent efficient fluorescence case. The smaller shaded circles illustrate a fainter fluorescence case equivalent to fluorescence with 50 per cent efficiency. For comparison we also plot the positions of fluorescent minerals (purple diamonds), or vegetation (green triangles) and the colours of planets in our own Solar system (labelled grey points). The changes in colour positions for the other stellar models are not significantly different.

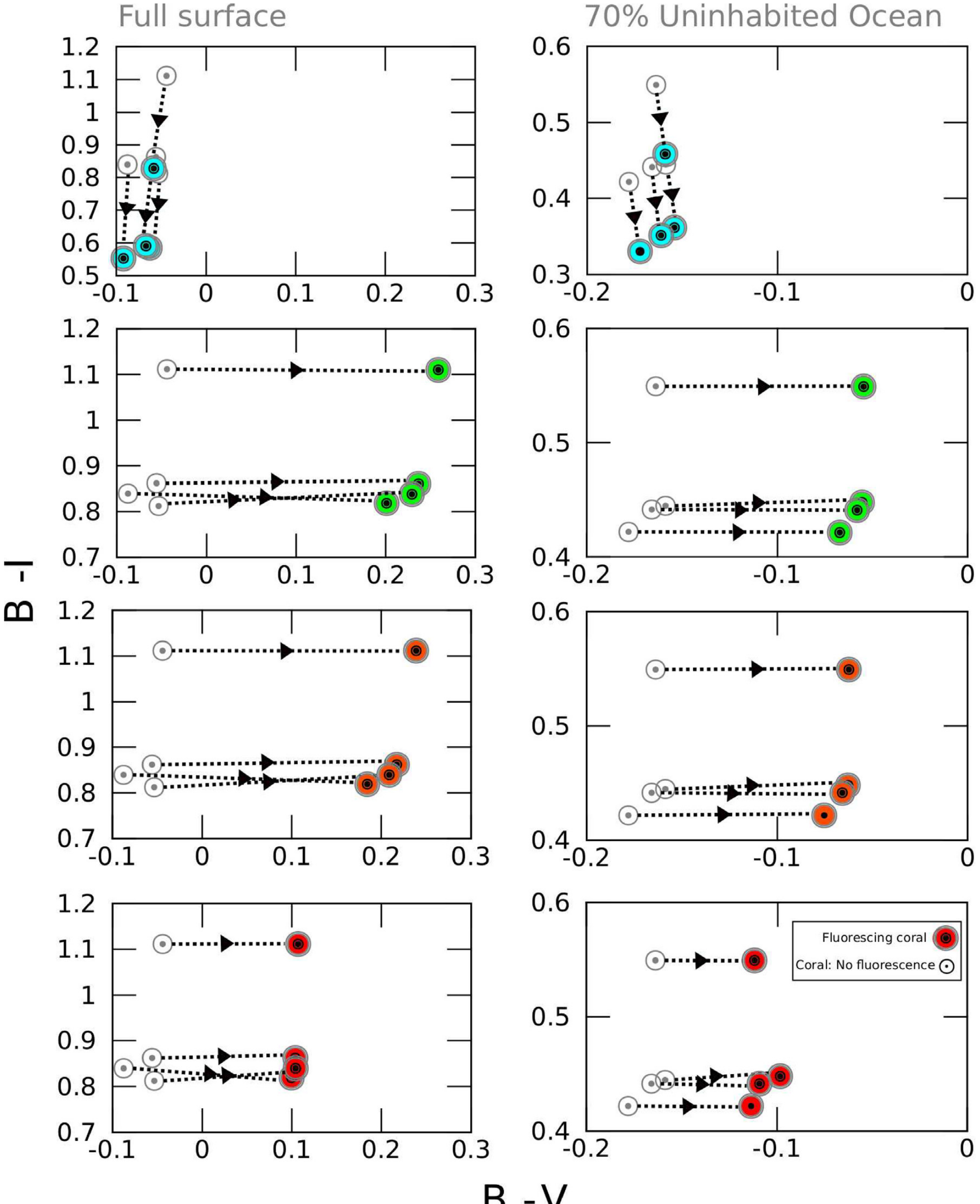

**Figure 7.** Colour–colour diagrams showing the surface colour of our planet models for atmospheres with 50 per cent cloud cover under an M0 star radiation environment. Fluorescing corals are indicated by the shaded circles [the colour of the shading represents the fluorescent wavelength described in a particular panel: cyan (486), green (515), orange (575), red (685)]. The direction of movement through the colour space as the surface transitions from a non-fluorescent to a fully fluorescent state is indicated by dashed lines. Here we show the case of bright, 100 per cent efficient fluorescence. Note the decrease in scale compared to Fig. 6, illustrating the reduction in magnitude of the temporal biosignature under cloudy conditions.

however, this specific case is included to show that such confusion can potentially occur if only one observation of such a biosphere is made during a UV flare.

Increasing cloud- and decreasing surface-coverage reduce the possibility of identifying biofluorescence as well as any other surface biosignatures. The presence of an Earth-like uninhabited ocean fraction does not affect the distinguishability of a temporal colour change due to biofluorescence (Fig. 6). The amount of colour change reduces with reducing surface fraction of the biofluorescent biosphere. The biofluorescent proteins at 515 and 575 nm show the widest spread in colour space during their fluorescent stage. The colours are distinct from these false positives in colour space, before and during fluorescence, at the four common fluorescent protein emission wavelengths (Fig. 6), except for the biofluorescent proteins at 515 and 575 nm for the fluorescent stage of Coral C and Coral D, which are very close to Mars' colour position. This makes observing the change in colour essential to distinguish these two fluorescent stages from a Mars-like object.

For an Earth-like cloud coverage of 50 per cent all biofluorescent colours are distinct from these false positives in colour space for both full surface and surface-ocean fraction, before and during fluorescence at the four common fluorescent protein emission wavelengths (Fig. 6). Increasing cloud cover moves the colours of all planetary models closer together, regardless of whether the planet is completely covered by minerals, vegetation, or a biofluorescent biosphere. Note that a planet model with 100 per cent cloud coverage would not show any surface feature and appear the same in spectra and colour, no matter what the underlying surface were covered in. The increase in cloud coverage also reduces the magnitude of the observable shift in position during biofluorescence, because only part of the surface is visible; the rest is blocked from our view by clouds (see Section 4).

**Table 3.** The properties of nearby M star systems that are currently known to harbour rocky habitable zone (HZ) planets. These represent our current best targets in searches for biosignatures, such as biofluorescence. Note that for the *TRAPPIST-1* system we include the outer planet (*TRAPPIST-1 h*), because this falls within the volcanic hydrogen HZ (Ramirez & Kaltenegger 2017). We calculate the angular separation [$\theta$ (arcsec) = $a$ (au)/$d$ (pc); where $a$ = planet semimajor axis, $d$ = distance to star system] for the HZ planets in each system. This is a useful quantity for inferring which instruments would be able to characterize these planets. The inner working angle (IWA) for a telescope describes the minimum angular separation at which a faint object can be detected around a bright star. For example, the planned 38 m E-ELT will have an IWA of ∼0.006 when observing in the visible region of the spectrum [assuming $\theta_{\rm IWA} \approx 2(\lambda/D)$, where $\lambda$ is the observing wavelength and D is the telescope diameter]. This makes *Proxima-b, Ross-128b*, and *LHS-1140b* particularly good observing targets for future characterization. Unless otherwise stated, stellar data were obtained from Anglada-Escudé et al. (2016) *[Proxima Cen.]*; Bonfils et al. (2018) *[Ross 128]*; Gillon et al. (2017) *[TRAPPIST-1]*; Dittmann et al. (2017) *[LHS 1140]*.

| | Proxima Centauri | Ross 128[a] | TRAPPIST-1 | LHS-1140 |
|---|---|---|---|---|
| *Type* | M5.5/6 V | M4 V | M8 V | M4.5 V |
| *Distance (pc)* | 1.3 | 3.38 ± 0.006 | 12.1 ± 0.4 | 12.5 ± 0.4 |
| *Luminosity (L)* | 0.001 55 ± 0.000 06 | 0.003 62 ± 0.000 39 | 0.000 524 ± 0.000 034 | 0.002 98 ± 0.000 21 |
| $T_{\rm eff}$ *(K)* | 3050 ± 100 | 3192 ± 60 | 2559 ± 50 | 3131 ± 100 |
| *Age (Gyr)* | 4.8[b] | ≥5 | 3–8[c] | >5 |
| *HZ (au)* | 0.03–0.09 | 0.05–0.13 | 0.02–0.05 | 0.04–0.11 |
| *HZ angular separation (milliarcsec)* | 23–69 mas | 14–38 mas | 1.6–4.1 mas | 3.2–8.8 mas |
| *Planet(s): Orbital distance (au), angular separation (milliarcsec)* | *Prox. Cen. b 0.0485 au:* 37 mas | *Ross 128b*[a] *0.05 AU: 15 mas* | *TRAPPIST-1e 0.028 au:* 2 mas<br>*TRAPPIST-1f at 0.037 au:* 3 mas<br>*TRAPPIST-1 g at 0.045 au:* 4 mas<br>*TRAPPIST-1 h at 0.06 au:* 5 mas | *LHS-1140b 0.0875 au:* 7 mas |

[a]Note that *Ross-128* was included here as an example, but is currently classified as an inactive star.
[b]Bazot et al. (2016).
[c]Luger et al. (2017b).

## 3.3 Detectability of biofluorescence on exoplanets

Biofluorescence can increase the visible flux at peak emission wavelengths temporarily by up to two orders of magnitude for a widespread biofluorescent biosphere, increasing the detectability of a planet due to the added emitted flux [similar arguments have been made for the detection of green light emission from aurorae on *Proxima-b* (Luger et al. 2017a)]. Biofluorescence would increase the overall visible flux, which is the sum of the reflected as well as emitted biofluorescence flux, improving the star–planet contrast ratio at these wavelengths. In an M star system, the reflected visible flux from a planet will be low due to the host star's low flux at these wavelengths; however, the proposed biofluorescent flux is based on the star's UV flux, and the resulting additional visible flux is therefore independent of the low stellar flux at visible wavelengths, increasing the visible light of a planet orbiting a cool active M star substantially (see Table 2).

Direct imaging of a HZ planet depends on the apparent angular separation of the planet from its host star as viewed from Earth. In Table 3, we list the currently known HZ planets orbiting the four closest M stars. For upcoming telescopes like the 38 m ELT with a proposed inner working angle at visible wavelengths of approximately 6 milliarcsec (mas) (Fujii et al. 2018); the orbital separation of the planets, *Proxima-b (37 mas), Ross 128b (15 mas)*, and *LHS-1140b* (7 mas) at maximum separation make these exoplanets prime candidates for direct imaging and the potential search for atmospheric as well as surface features indicating life (see Table 3).

### 3.3.1 Earth-analogue planets

To allow easy comparison, we model the contrast ratio increase for a 1 Earth radius and an Earth-analogue planet instead of modelling the unknown exoplanets in detail. The size of the exoplanets will influence the results with an increased planetary radius and surface area increasing the overall flux, and a decreasing surface coverage and increased cloud coverage decreasing the ratio. We show that the star–planet contrast ratio in the visible regime induced by fluorescence can decrease by up to two orders of magnitude from $\sim10^{-8}$ to $\sim10^{-6}$ at peak emission wavelengths. This would bring them within the range of the capabilities of instruments with adaptive optics planned for the E-ELT (Turbet et al. 2016), as well as planned future space-based coronagraph missions [which potentially could achieve even higher contrast ratios (see e.g. Traub & Oppenheimer 2010)]. Higher contrast ratios are needed to potentially detect fainter global fluorescence, fluorescence with a more limited UV absorption range, or fluorescence on planets with only partially inhabited surfaces. Similar HZ planets around bright nearby M stars are likely to be uncovered by missions like TESS (Ricker et al. 2014; Barnes et al. 2015; Sullivan et al. 2015), adding to the list of potential fluorescent biosignature search targets. Note that the colour–colour diagram is a relative measurement and thus is not influenced by the absolute flux from a planet.

We are now in an era where we are building up catalogues of potential targets to search for biosignatures. Results from the first TESS data release suggest potential criteria for selecting the best host star targets (Kaltenegger et al. 2019) for biofluorescent biosignatures. Günther et al. (2019) show that mid-M type stars are the most numerous and have the highest fraction of flaring stars. A target list could be further narrowed down by selecting those that host a planet that exists within the overlap between the abiogenesis zone (Rimmer et al. 2018) – the zone in which UV flux would be sufficient to drive prebiotic chemistry – and the liquid water HZ. Günther et al. (2019) concludes that 28 M stars in the sample could have the required UV flux for prebiotic chemistry to occur on a HZ planet, although with the potentially negative consequence for surface habitability of eroded ozone layers. However, although it remains to be determined how high UV-B and -C fluxes would effect biofluorescent molecules (see section 4.3), a lack of an ozone layer on such planets may in fact be a positive factor in the search for the emergence of life as well as potential biofluorescent biosignatures. Hence, an aspect of M star habitability, high-flare activity, which is normally considered to be negative, should be considered in a new light. As more data are gathered on TESS stars and their exoplanets, a list of potential exoplanet targets could be further defined using information on host star flare energies and frequencies in combination with the orbital distance of rocky habitable planets.

# 4 DISCUSSION

## 4.1 The effect of clouds

The amount of cloud coverage cannot be self-consistently modelled for extrasolar planets for now. Therefore we show clear-sky and Earth-like cloud coverage to explore two points from a wide range of models. With decreasing surface pressure, cloud coverage should also decrease. As low-density atmospheres are more likely around active M stars, this could improve the detectability of any surface spectral biosignatures. In addition, low-density atmospheres would provide a higher surface UV environment, increasing the emitted flux due to biofluorescence if it evolved, thus our results are conservative estimates and should increase for lower density atmospheres or atmospheres without oxygen.

To remove the signal due to clouds and assess any surface features for planets with non-complete cloud coverage the signal due to reflection from clouds can be distinguished from surface features with many short, high signal-to-noise observations, because clouds should occupy all areas of the planet given enough time. Thus one can separate them from surface features that are bound to the rotation of a planet, if the observations can be limited to about 1/20 of the planet's rotation period, or for the Earth, about an hour (see Pallé et al. 2008 for details), requiring big future telescopes to allow such time-resolved observations.

## 4.2 Duration of elevated UV flux

Our results illustrate the maximum strength the biofluorescent temporal biosignature could reach. A temporal signature such as this will vary as the surface UV flux on a planet's surface varies, increasing in strength alongside the flare, reaching a maximum alongside the flare peak, then decrease as the flare event subsides. Using the AD Leo flare as a guide, this duration of a biofluorescent signal would last several hours, similar to the total flare duration. However, the maximum UV surface flux associated with the flare peak would only last for approximately 15 min. Using the values given for the AD Leo flare in Segura et al. (2010), the time-averaged UV-A flux would be approximately an order of magnitude lower than its peak values, which, for 100 per cent efficient biofluorescence, would result in a order of magnitude lower increase in visible flux, compared to those we predict in Table 2. This would in particular make the fractional surface coverage scenarios more challenge to detect. However, this assumes an isolated flare event. Particularly active M stars, such as Proxima Centauri, flare very frequently (several times a day). This could result in longer periods of elevated UV surface fluxes, suggesting that the best targets for a biofluorescent biosignature would the most active M stars in a particular sample.

## 4.3 Evolutionary arguments

In this hypothesis we use corals as an example of an organism that exhibits fluorescence and contains convergent evolutionary features (i.e. features that have evolved independently in different species that are not closely related). Convergent evolution arguments often underpin the definition of plausible astrobiological biosignatures (see e.g. Chela-Flores 2007; Martinez 2015; Cockell 2018). In a high UV surface environment, the evolution of an organism that shares some of these convergent traits with corals – specifically the ability to form simple, collaborative colonies and the ability to fluoresce – should be beneficial, allowing life to take advantage of surface environments, e.g. to gain nutrition and energy from visible light.

Corals were one of the earliest forms of animal life to evolve on Earth, with an evolutionary history spanning 500 Myr. Initially existing as simple, solitary organisms, they later evolved into collective reefs. Reef-building is an ancient trait for life on Earth. Many marine species have independently evolved reef-building abilities, but the trait can be traced back to colonial groups of bacteria building stromatolites 3.5 Gya (Kiessling 2009).

The fluorescent pigments and proteins in corals are descended from GFPs (Field et al. 2006). GFPs are present in a variety of phyla, suggesting an origin within an ancient common ancestor from amongst very early metazoan (animal) life, over 500 Myr ago (Chudakov et al. 2010). Biofluorescence is widespread in life on Earth and is thought to have evolved independently, multiple times (Sparks et al., 2014), which strengthens a case for the evolution of biofluorescence on other inhabited worlds, following convergent evolution arguments. Furthermore, in a study on coral fluorescent proteins, Roth et al. (2010) found that rapid changes in protein concentrations occurred in response to changes in light intensity, and that pigment concentration strongly correlates with fluorescence intensity. This suggests that the evolution of dense fluorescent pigment concentrations in a high-UV environment may be favoured.

The ozone layer limits Earth's surface UV flux to UV-A wavelengths, so fluorescent molecules on Earth only absorb photons from this part of the UV spectrum. On the M star planets we consider here, significant UV-B and UV-C fluxes could reach the surface (compared to Earth). For example, during a flare, models show that a planet in the HZ of *AD Leo* would receive a UV-B surface flux 80 per cent higher than Earth's (Segura et al. 2010), while the UV-C surface flux would be several orders of magnitude higher than Earth's, if an ionized particle stream aligns with the flare and depletes the planet's ozone layer (Segura et al. 2010; Tilley et al. 2019). Furthermore, protective ozone layers may not be present in the atmospheres of habitable exoplanets. Atmospheres that resemble that of the early Earth prior to the great oxidation event (GOE) would allow the majority of the top of the atmosphere UV flux to reach the surface. For habitable planets in active M star systems this leads to UV-C and -B fluxes that are several order of magnitude higher than they would be if a protective ozone layer were present; though notably these levels still be at least an order of magnitude lower than levels on the pre-GOE Earth (O'Malley-James & Kaltenegger 2018b).

Hence, fluorescent proteins that downshift the full UV flux (UV-A to -C) to benign visible wavelengths (perhaps via a cascading chain of fluorescent proteins with different excitation wavelengths; see for example, Gilmore et al. 2003) may be selected for under high-UV conditions, further strengthening a biofluorescent signal during flares. From a biosignature detectability perspective, the downshifting of a much wider range of UV flux could result in stronger biofluorescent signals in the visible than those that we estimate here, using the known behaviour of coral fluorescent pigments and proteins that absorb UV-A radiation. Estimating the nature of a biofluorescent signal resulting from the downshifting of the full UV-A to -C flux is challenging given that there are no biological analogues on present-day Earth that do this. Thus we do not include such organisms into our analysis, but we discuss challenges for such organisms shortly in the Section 4. An organism that evolved to do this would need to overcome challenges. This raises many questions about how an organism might do this. For example, the peak absorption of the DNA molecule (and many other biomolecules) occurs in the UV-C; hence, simply shifting the absorption peak into that range could cause more biological harm. Where would the optimum absorption peak be for biofluorescence under these circumstances? If UV radiation is stepped down over a much wider range, would this also influence the efficiencies the fluorescence process could achieve? We lack terrestrial biological analogues from which we could glean potential answers to questions like these. This highlights interesting experimental avenues to explore the responses of terrestrial fluorescent proteins and pigments to high and temporarily high (mimicking flare events) UV-C and -B environments, similar to those expected on planets in M star systems.

There is some evidence from Mars exposure simulations that autofluorescence signal from photosynthetic pigments can still occur after exposure to high UV-B and -C levels similar to those on the surface of Mars, but at much degraded level of 35 per cent after 64 h (Dartnell & Patel 2014). This hints at the possibility of fluorescent organisms adapting and optimizing for such periodic high-UV exposures over time and suggests an avenue of experimental research to further understand such possibilities.

### 4.4 Comparison to other surface biosignatures

We compare our models to the only other widespread surface biofeature on Earth, vegetation, and its proposed surface biosignature, the vegetation red edge (VRE) (e.g. Sagan et al. 1993; Seager et al. 2005; O'Malley-James & Kaltenegger 2018a). Vegetation on Earth creates a weak spectral signature that causes an increase in reflectance of a few to 10 per cent in the NIR (see review by Arnold 2008 and references therein) for present-day Earth. As we show in Fig. 5, biofluorescence signals can be stronger than a VRE signal under similar conditions (surface fractions, cloud coverage). Therefore, if strong biofluorescence evolves on a planet, it could produce a stronger signal than vegetation on Earth. If the fluorescent proteins used by life in our model biosphere did not evolve to have higher efficiencies due to a high-UV environment, but were restricted to the common Earth-like efficiencies of coral fluorescent proteins ($\sim$10 per cent), the change in global brightness at the peak wavelength of fluorescence only increases by a factor of 2 (clear skies) and $\sim$1.5 (50 per cent cloud cover). Note that any surface signatures will be challenging to detect on exoplanets, especially with large cloud coverage.

## 5 CONCLUSIONS

In this work, we explored the hypothesis and detectability of exoplanets dominated by a biofluorescent biosphere that uses fluorescence as a UV damage-mitigation strategy. Planets in the HZ of active M stars, like *Proxima-b*, would be exposed to temporal high ultraviolet (UV) radiation. During an M star flare, the UV radiation flux on the surface of a HZ planet can increase by up to two orders of magnitude. Photoprotective biofluorescence (the 'up-shifting' of UV light to longer, safer wavelengths, via absorption by fluorescent proteins), a proposed UV protection mechanism in some coral species, would not only mitigate UV radiation effects on an organism, but could also increase the detectability of such biota in the visible range of the spectrum due to additional emitted visible flux. Such biofluorescence could be observable as a 'temporal biosignature' for planets orbiting stars with changing UV environments, like active flaring M stars. Depending on the efficiency of the fluorescence, biofluorescence can increase the visible flux of a planet at the peak emission wavelength by over an order of magnitude during a flare event. For comparison, the change in brightness at peak emission wavelengths caused by biofluorescence could increase the visible flux of an Earth-like planet by two orders of magnitude for a widespread biofluorescent biosphere and clear skies, with low-cloud scenarios being more likely for eroded atmospheres. In an M star system, the reflected visible flux from a planet will be low due to the host star's low flux at these wavelengths; however, the proposed biofluorescent flux is dependent on the host star's UV flux, resulting in additional visible flux that is independent of the low stellar flux at visible wavelengths. This suggests that exoplanets in the HZ of active M stars are interesting targets in the search for signs of life beyond Earth.

Using a standard astronomy tool to characterize stellar objects, a colour–colour diagram, one can initially distinguish planets with and without biofluorescent biosignatures due to the temporal shift in their colours. The extent of the colour change depends on the efficiency of the fluorescence, surface fraction covered and cloud coverage. The change in colour caused by biofluorescence differs, in position and magnitude, from that caused by abiotic fluorescence, distinguishing both. Biofluorescence is separated in traditional astronomical colour–colour space from other surface biosignatures, known Earth-based fluorescent minerals and other Solar system planet colours. In colour–colour space there is no conflict (i.e. false positives) with Solar system planets, except for a 100 per cent fluorescence clear-sky scenario for the colour of Mars. Increasing cloud- and decreasing surface-coverage not only moves the colours of our model planets away from Solar system planet colours, but also reduce the possibility of identifying biofluorescence as well as any other surface biosignatures.

Exoplanets around close-by active M stars are promising targets to search for such brightness changes and spectra indicating such a surface biosphere for upcoming ELTs. Especially the recently discovered planet *Proxima-b*, in the HZ of *Prox. Cen*., an active M star, as well as *LHS-1140b* are future candidates for such direct observations. At maximum orbital separation from their host star, these planets can be resolved with upcoming telescopes like the ELT. Future discoveries by ground based, as well as the TESS mission, surveys of HZ planets around bright, active nearby M stars will provide us with additional observation targets for the search for fluorescent biospheres. If biofluorescence can evolve on planets orbiting active M stars, high-UV fluxes, rather than being deleterious for the detection of life, could make certain hidden biospheres detectable.

## ACKNOWLEDGEMENTS

The authors acknowledge helpful discussion with Charles Mazel, Juan Torres-Pérez for providing coral spectral data, Sarah Rugheimer for providing atmosphere spectra and the USGS Digital Spectral Library and the ASTER Spectral Library for reflectance spectral data for minerals, vegetation, and ocean water. We thank an anonymous referee for comments that helped improve this manuscript. We would also like to acknowledge funding from the Simons Foundation and the Carl Sagan Institute at Cornell (290357, Kaltenegger).